\documentstyle[twocolumn,prc,aps]{revtex}
\input epsf
\begin{document}
\twocolumn[\hsize\textwidth\columnwidth\hsize\csname @twocolumnfalse\endcsname 
\title{Extraordinary Baryon Fluctuations and the QCD Tricritical Point}

\author{Sean Gavin}
\address{
Department of Physics and Astronomy, Wayne State University, 
Detroit, MI, 48202}
\date{\today} 
\maketitle
\begin{abstract}
  The dynamic separation into phases of high and low baryon density in
  a heavy ion collision can enhance fluctuations of the net rapidity
  density of baryons compared to model expectations.  We show how
  these fluctuations arise and how they can survive through freezeout.

\vspace{0.1in}
\end{abstract}
]

\begin{narrowtext}

QCD can exhibit a first order phase transition at high temperature
and baryon density, culminating in a tricritical point
\cite{BergesRajagopal,Steph1}. Specifically, below the tricritical
point, a phase coexistence region separates distinct phases of QCD
matter at different baryon densities, as shown in fig.~1.  Stephanov,
Rajagopal and Shuryak have pointed out that critical fluctuations of
$E_T$ and similar meson measurements can lead to striking signals at
the tricritical point in relativistic heavy ion collisions
\cite{Steph2}.

We suggest that measurements of fluctuations of the net baryon
number in nuclear collisions can help establish the first
order coexistence region and, ultimately, the tricritical point.  We
characterize baryon fluctuations by the variance ${\cal V}_B \equiv
\langle N_B^2\rangle - \langle N_B\rangle^2$, where $N_B = N -
\overline{N}$ is the net baryon number in one unit of rapidity,
obtained from the baryon $N$ and antibaryon $\overline N$
distributions; the average is over events. Ordinarily, net baryon
fluctuations in thermal and participant nucleon models of heavy ion
collisions satisfy
\begin{equation}
{\cal V}_{B}^0
\approx TV\partial \rho_B/\partial \mu_B \approx N + \overline{N},
\label{eq:i1}\end{equation}
where the second equality holds for an ideal gas \cite{gp1,gp2}.  In
contrast, we argue below that enhanced fluctuations occur if the
expansion of the system quenches the matter from an initial high
density state into the phase coexistence region.
At the tricritical point, these fluctuations {\em diverge} because
$V\partial \rho_B/\partial \mu_B = -\partial^2 \Omega/\partial\mu_B^2
\rightarrow \infty$ in (\ref{eq:i1}), where $\Omega$ is the free energy.

Baryon fluctuation measurements add new leverage to a search for the
first order region, complimenting information from pion and kaon
interferometry, intermittency and wavelet
analyses \cite{Pratt}. The latter measurements probe the spatial
structure introduced by phase separation and droplet formation.  By
comparison, baryon fluctuations are weakly dependent on the morphology of
the mixed phase, because they do not rely on distinct droplets
escaping a rather dense system. 

In this paper we explore the onset of baryon fluctuations and the
possibility of their dissipation in nuclear collisions.  That an order
parameter such as the baryon density should undergo extraordinary
fluctuations during a phase transition comes as no surprise --
critical opalescence results from an analogous divergence of density
fluctuations.  Furthermore, less extreme but nevertheless measurable density
fluctuations are familiar in condensed matter systems that are rapidly
quenched into a phase coexistence region \cite{spinodal}.  Perhaps
more surprising is the possibility that observable baryon fluctuations
can survive the subsequent evolution of the system.

As motivation, we start by describing how phase separation can produce
observable fluctuations.  We then use a spinodal decomposition model
\cite{spinodal} to illustrate how a highly supercooled system can
produce large fluctuations.  Finally, we ask whether diffusion can
dissipate these fluctuations before they are detected.  Generally,
there are two ways the dynamics can obscure large net baryon
fluctuations in a subregion of the system. First, particles can
diffuse throughout the fluid, diluting any ``hot spots'' and their
consequent fluctuations.  Second, fluid flow can carry the particles
away to similar effect.  We focus on the effect of diffusion because
transport theory estimates indicate that the relevant diffusion
coefficient can be large \cite{PVW,HeiselbergPethick}. In principle,
chemical reactions introduce dissipation by annihilating and creating
baryon-antibaryon pairs.  However, these reactions cannot affect the
net baryon number, which is conserved, although they do affect the
individual baryon and antibaryon fluctuations \cite{gp2}.
\begin{figure} 
\epsfxsize=2.5in
\centerline{\epsffile{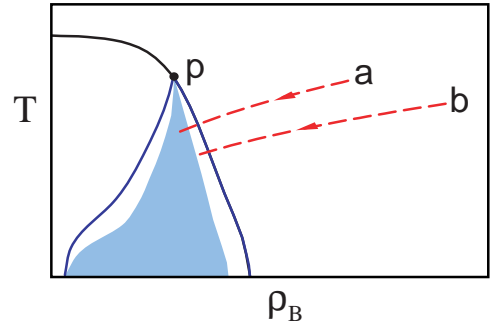}}
\vskip +0.05in
\caption[]{
  Schematic QCD phase diagram \cite{BergesRajagopal,Steph1} including
  the tricritical point (p) and the spinodal region (shaded).  The
  dashed curves indicate trajectories followed by ion collisions.

}\end{figure}

Figure~1 shows the expected QCD phase diagram at high temperature $T$
and baryon density $\rho_B$.  A phase coexistence region lies below
the tricritical point. Inside this region, uniform matter must
eventually break up into distinct domains containing the high and low
density phases respectively. Within the coexistence region is the
spinodal region (the shaded area). There, uniform matter is
mechanically unstable, so that small fluctuations can rapidly generate
bubbles throughout the fluid. Matter is metastable between the
spinodal and coexistence boundaries, so that fluctuations must
overcome an energy barrier to nucleate bubbles.  Phase separation by
spinodal decomposition and bubble nucleation \cite{spinodal} have been
discussed in the context of heavy ion collisions
\cite{bubbleHI,supercool}, albeit with different phase diagrams in
mind.

Ideally, a heavy ion collision will produce extraordinary fluctuations
if experimenters can adjust the beam energy and the ion combination to
produce initial baryon densities and temperatures within the spinodal
region or, optimally, at the tricritical point.  Alternatively,
fluctuations can arise if the expansion of the heavy ion system
rapidly quenches a high density system deeply into the spinodal
region. The dashed curves in fig.~1 show one such quenching trajectory
(a) and one trajectory that only reaches the nucleation region (b).
Either process may produce phase separation far from equilibrium
\cite{supercool}.

In spinodal decomposition, runaway density fluctuations rapidly
contort and compress the high density fluid into filamentary regions.
These runaway modes appear because the fluid is dynamically unstable,
with $\partial \rho_B/\partial \mu_B < 0$ inside the spinodal region
\cite{spinodal}.  These modes grow exponentially until the density
outside the regions reaches the equilibrium density for that
temperature, $\rho_h$, at the low-density boundary of the phase
coexistence curve. After this burst of nonequilbrium growth, phase
separation in an ion collision can proceed smoothly as the system
expands and rarefies.  Nucleation may proceed more uniformly, with
perhaps one bubble growing smoothly as the system expands.
However, there is little distinction between spinodal decomposition and 
nucleation in a highly supercooled system \cite{supercool}.

To understand why these baryon density fluctuations can exceed our
baseline estimate (\ref{eq:i1}), suppose that by a proper time
$\tau_Q$ the phase transition has created a relatively stable initial
fraction $1-f$ of the low density bubbles within the high density
phase. The net baryon density is then $\rho_B = f\rho_q +
(1-f)\rho_h$, where $\rho_q$ and $\rho_h$ are the equilibrium
densities of the respective phases.  This density corresponds to a net
rapidity density of baryons $N_B \approx {\cal A}\rho_B\tau_{Q}$,
where ${\cal A}$ is the transverse area of the collision volume.  We
write
\begin{equation}
N_B \approx f N_q + (1-f) N_h,
\label{eq:mean}\end{equation}
where $N_{q,h} \equiv \rho_{q,h}{\cal A}\tau_{Q}$.  The fluctuations
of $N_B$ are enhanced relative to a uniform system because the
distribution of densities within each event is bimodal, with peaks at
$\rho_q$ and $\rho_h$. The variance is therefore
\begin{equation}
{\cal V}_B \approx {\cal V}_{B}^0 +
f(1-f)(N_q-N_h)^2,
\label{eq:fluct}\end{equation}
where ${\cal V}_{B}^0 = f{\cal V}_q + (1-f){\cal V}_h$ is the weighted
average of the variance of each phase.  If we take each component to
be a nearly ideal gas, then ${\cal V}_{q,h} = N_{q,h} +
\overline{N}_{q,h}$, where $N$ and $\overline{N}$ are the rapidity
densities of baryons and antibaryons \cite{gp1}. The quantity ${\cal
  V}_{B}^0$ then reduces to (\ref{eq:i1}), precisely the Poissonian
fluctuations we would expect if the system were uniform. The total
variance (\ref{eq:fluct}) exceeds this value by an amount proportional
to the square of the density contrast of the two phases. A
nonequilibrium model of the quench would be needed to compute $f$.
Observe that the two-phase effect, which vanishes with the density  
contrast, gives way to the critical divergence of
(\ref{eq:i1}) near tricritical point.
  
To estimate the affect of flow on baryon evolution, we write the net
baryon current conservation law:
\begin{equation}
\partial\rho_B/\partial \tau  + \partial_\mu j_B^\mu = 0. 
\label{eq:eq8}\end{equation}
where $\partial/\partial\tau \equiv u^\mu\partial_\mu$ for a fluid of
four velocity $u^\mu$. The flow of the system changes the baryon
density through $u^\mu$. If we neglect dissipation for the moment, then
the net baryon number flows along with fluid as a whole, i.e.
$j_B^\mu = \rho_B u^\mu$.  For the Bjorken scaling flow, $u^\mu =
u_{\rm s}^\mu \equiv \tau^{-1}(t, 0, 0, z)$ and $\tau \approx
\sqrt{t^2 - z^2}$. As is well known, the density satisfies
$\rho_B(\tau)\tau = \rho_B(\tau_Q)\tau_Q$ for scaling flow.  The
rapidity density is then $\tau$ independent.  The variance of $N_B$
for an ensemble of events $i$ is ${\cal V}_B = \sum_i (N_B^i -
N_B)^2$.  Differentiating, we find the scaling results
\begin{equation}
dN_B/d\tau \approx 0\,\,\,\,\,\,\,\,{\rm and}\,\,\,\,\,\,\,\,
d{\cal V}_B/d\tau \approx 0.
\label{eq:rap}\end{equation}
Even though the expansion after $\tau_Q$ will cause $f(\tau)$ to
decrease, both $N_B$ and ${\cal V}_B$ are fixed at the initial values
(\ref{eq:mean}, \ref{eq:fluct}). It follows that (\ref{eq:fluct}) can
represent the {\em observed} fluctuations, provided that the flow
satisfies scaling. 

Baryon diffusion can play an important role in both the onset and the
propagation of fluctuations. We follow \cite{LL} and
write
$j_B^\mu = \rho_B u^\mu + j_{\rm diss}^{\mu}$, 
where we define $u^{\mu}$ so that the total momentum density of the
fluid vanishes in the local rest frame.  The diffusion of baryons
relative to the fluid center of momentum gives rise to:
\begin{equation}
j_{\rm diss}^\mu = -MT \nabla^\mu \left({{\mu_B}/{T}}\right),
\label{eq:eq11}\end{equation}
where $M = D\partial\rho_B/\partial \mu_B$ is loosely termed the
mobility, $D$ is the diffusion coefficient and $\nabla^\mu =
(g^{\mu\nu} + u^\mu u^\nu)\partial_\mu$.  

To illustrate the onset of spinodal decomposition, we modify
(\ref{eq:eq11}) to describe the strong inhomogeneities that
spontaneously arise in the fluid. We follow the classic linear
stability analysis of Cahn \cite{CahnHilliard}, duplicating the
salient details here {\it a}) to motivate QCD calculations of the
microscopic inputs and {\it b}) to highlight the differences and
similarities with DCC studies \cite{dcc}. We study the evolution of a small
perturbation ${\tilde\rho}_B(k,t) = {\tilde\rho}_B(t)\exp(i{\vec
  k}\cdot{\vec r})$ in the spinodal region, where
$\partial\rho_B/\partial \mu_B$ is negative.  Anticipating that the
fluid will spontaneously become inhomogeneous, we assume that the free
energy of ref.~\cite{BergesRajagopal} can be written as an effective
Ginzburg-Landau functional of the baryon density,
\begin{equation}
F\{\rho_B(\vec{r})\}
= \int d^3r\{f[\rho_B(\vec{r})]+\xi(\nabla \rho_B)^2/2\},
\label{eq:sp1}\end{equation}
in the local rest frame, where $f=\Omega/V + \mu_{B} \rho_{B}$ is the Helmholtz 
free energy density for a uniform system.  Replacing $\mu_B = 
(\partial F/\partial N_B)_{T,V}$ in
(\ref{eq:eq11}) with the functional derivative with respect to
$\rho_B$, we find that
\begin{equation}
{{\partial \over{\partial t}}{\tilde\rho}_B(t)} = 
-{M k^2}\left({{\partial\mu_B}\over{\partial \rho_B}} + 
{{\xi}\over{M}} k^2\right) {\tilde\rho}_B(t).
\label{eq:sp2}\end{equation}
For $\partial\rho_B/\partial \mu_B<0$, a perturbation of 
$k < k_{\rm c} =[|\partial\rho_B/\partial\mu_B|/M\xi]^{1/2}$
grows exponentially. The fastest mode at $k_{\rm sp} = k_{\rm c}/\sqrt{2}$ 
grows at the shortest time scale 
\begin{equation}
\tau_{\rm sp} = {{2|\partial\rho_B/\partial \mu_B|}\over {Mk_{\rm sp}^2}},
\,\,\,\,\,\,\,\,{\rm where}\,\,\,\,\,\,\,\, 
k_{\rm sp} = {{k_{\rm c}\over {\sqrt{2}}}}.
\label{eq:sp3}\end{equation}
This mode dominates the early stage of the decomposition.  Outside the
spinodal region where $\partial\rho_B/\partial \mu_B > 0$,
perturbations decay exponentially at rates that differ from normal
diffusion only for wavelengths smaller than $2\pi/k_{\rm c}$ (roughly
the equilibrium droplet size in a vacuum).  At these early times, the
variance averaged over the thermal fluctuations within an event grows
as $\langle\langle \tilde{{\rho}}_B(r,t)^2\rangle\rangle \propto {\rm
  e}^{2t/\tau_{\rm sp}}$.  This growth exceeds that of the
density, $\propto {\rm e}^{t/\tau_{\rm sp}}$, allowing large
fluctuations to build.

A quantitative treatment of spinodal decomposition using modern
techniques \cite{spinodal} awaits the introduction of kinetic terms to
the coarse-grained free energy density \cite{BergesRajagopal,Steph1}
--- neither $M$ nor $\xi$ are known.  However, we can get a rough idea
of the time scales needed for the onset of instabilities by
identifying the inverse momentum $k_{\rm sp}$ with the correlation
length, $m_\sigma^{-1}\sim 1$~fm, as in \cite{Steph2}. Well within the
spinodal region, we surmise that
$|M(\partial\rho_B/\partial\mu_B)^{-1}|$ is of the order of the
diffusion coefficient in a stable, perturbative plasma, $ \sim 1-3$~fm
\cite{HeiselbergPethick}.  Therefore $\tau_{\rm sp}\sim 1$~fm or
smaller, suggesting that spinodal decomposition can erupt violently
when a quench is achieved.  

One faces a similar situation in disoriented chiral condensate
formation \cite{dcc}. There, one studies nonequilibrium fluctuations
of the chiral order parameter at very small baryon density.
Equilibrium fluctuations of that order parameter diverge along with the 
baryon density fluctuations at the tricritical point 
\cite{BergesRajagopal}.  The essential distinction between our first order 
transition and those phenomena is that, here, the low density phase can 
coexist with the high density phase in thermodynamically stable bubbles.  
This is impossible in a second order transition.  The wavelength scale 
$\sim 2\pi/k_{\rm sp}$ of the fastest growing mode (\ref{eq:sp3}) is a 
nonequilibrium manifestation of the equilibrium size of bubbles.  The 
second order chiral transition is also heralded by unstable 
modes, but the fastest mode has $k=0$, reflecting the critical divergence 
of the correlation length.  Furthermore, the corresponding time scale 
diverges in a ``critical slowing down'' as the correlation length $\sim 
m_\sigma^{-1}\rightarrow \infty$, see, e.g., eq.  (2) in \cite{ggp}.  
Analogous behavior takes place at the tricritical point, where 
$M^{-1}\partial\rho_B/\partial\mu_B$ and (\ref{eq:sp3}) diverge.

Optimistic that fluctuations can be produced by a quench, we turn to
the dissipation of these fluctuations due to diffusion. 
For a stable liquid, we use (\ref{eq:eq8},\ref{eq:eq11}) to obtain 
the diffusion equation,
\begin{equation}
{{\partial \rho_B}/{\partial \tau}} + \rho_B\partial_\mu u^\mu = 
D \nabla^2 \rho_B,
\label{eq:eq11c}\end{equation}
where we take $D$ to be constant and neglect thermodiffusion.
Observe that diffusion {\em cannot} affect scaling flow because the
diffusive term vanishes if $\rho_B$ is a function of $\tau$ alone. In
that case (\ref{eq:rap}) would still apply. Let us then face a 
conservative scenario in which phase separation only occurs as a 
perturbation $\tilde{\rho}(\tau,\eta)$ of the scaling
hadronic system (here $\tanh\eta = z/t$ is the spatial rapidity). To
be concrete, we take $N_B$ and ${\cal V}_B$ to be perturbed
at time $\tau_Q$ by $\tilde{N}_B \sim f(N_q -N_h)$ and $\tilde{{\cal
    V}}_B \sim f(1-f)(N_q -N_h)^2$ for 
$f = f_0\exp\{-\eta^2/2{\sigma}^2\}$. Note that $\sigma \approx 0.88$ for
a spherically symmetric perturbation.

We describe the evolution for $\tau > \tau_Q$ using (\ref{eq:eq11c})
for $\rho_B \partial_\mu u^\mu \approx \rho_B/\tau$. Writing the
rapidity density at fixed $\tau$ as $N_B(\tau, \eta) =
\rho_B(\tau,\eta){\cal A}\tau$, we obtain
\begin{equation}
\tau^2\partial \tilde{N}_B /\partial \tau = D\partial^2  
 \tilde{N}_B/\partial \eta^2.
\label{eq:eq12}\end{equation}
The unperturbed rapidity density $N_B^0$ is constant. It follows that
${\tilde N}_B = \tilde{N}_B(\tau_Q)\phi_\sigma(\eta)$, 
where $\tilde{N}_B(\tau_Q) = f_0(N_q - N_h)$ and
\begin{equation}
\phi_\sigma = {{\exp\{-\eta^2/2({\sigma}^2 + 2Ds)\}}\over
{(1+2Ds\sigma^{-2})^{1/2}}},\,\,\,\,\,\,\,\,\, 
s = \tau_Q^{-1} - \tau_F^{-1},
\label{eq:eq13b}\end{equation}
and $\tau_F$ is the freezeout time. Longitudinal flow limits the
degree to which the Gaussian perturbation can be dispersed.  For
$\tau_F \gg \tau_Q$, we see that the rapidity density near $y \approx
\eta \approx 0$ is $N_B \approx N_B^0 + \tilde{N}_B(\tau_Q) \{1 +
2D/(\tau_Q\sigma^2)\}^{-1/2}$ since $s\approx\tau_Q^{-1}$.

To study the contribution of the perturbation to the variance for
$\tau \gg \tau_Q$, we write the event averaged ${\cal V}_B \approx
{\cal V}_{B}^0 + 2\langle N_B^{0}\tilde{N}_B\rangle$ \cite{linear}. We
then differentiate ${\cal V}_B$ with respect to $\tau$ and use
(\ref{eq:eq12}) to obtain:
\begin{equation}
\tau^2\partial \tilde{{\cal V}}_B/\partial \tau 
= D\partial^2 \tilde{{\cal V}}_B 
/\partial \eta^2,
\label{eq:eq14}\end{equation}
to linear order in $\tilde{N}_B/N_B$. We find
${\tilde{\cal V}}_B = (N_q-N_h)^2\{f_0\phi_\sigma(\eta)
-f_0^2\phi_{\sigma/\sqrt{2}}(\eta)\}$,
for $\phi_\sigma(\eta)$ given by (\ref{eq:eq13b}).

The diffusion coefficient for baryon current is unknown for the mixed
phase described in \cite{BergesRajagopal,Steph1}. However, we can get
a rough estimate for $D$ in the pure phases from kinetic theory, which
implies that $D\sim \tau_{\rm diff} \,v_{\rm th}^2/3$,
where $v_{\rm th}$ is the thermal velocity of baryons and $\tau_{\rm
  diff}$ is the relaxation time for diffusion. For nucleons diffusing
through a hadron gas, $\tau_{\rm diff}\sim 35$~fm and $D\sim 6$~fm
\cite{PVW} at a temperature of 150 MeV. Flavor diffusion through a
perturbative quark gluon plasma yields $D\sim 1-3$~fm
\cite{HeiselbergPethick}.
We estimate diffusion in the mixed phase using the larger hadronic
value, $D\sim 6$~fm. In fact, arguments in
ref.~\cite{BergesRajagopal} suggest similarities between hadrons and
droplets of mixed phase.  At $\eta\approx y\approx 0$, we find the rather
small decrease $\tilde{{\cal V}}_B/\tilde{{\cal V}}_B(\tau_{Q})\sim 76\%$
for $\tau_Q \approx 5$~fm, $\sigma = 0.88$, $\tau_F \rightarrow \infty$ 
and $f_0 = 1/2$. 

Experimenters can use baryon fluctuations to search for the tricritical
point as follows. Collisions for a range of beam energies, ion
combinations and centralities can produce high density systems that follow
trajectories as in fig.~1. Results can be compared as shown in fig.~2 by 
plotting the normalized ratio \cite{BaymHeiselberg},
\begin{equation}
\omega_B \equiv {\cal V}_B/(N + \overline{N})
\label{eq:eq17}\end{equation}
as a function of a normalized centrality selector, e.g. the total
charged particle multiplicity $N_{\rm ch}$. In the absence of unusual
fluctuations, this ratio is energy independent, with a value close to
unity, cf.~eq.~(\ref{eq:i1}). In fig.~2 we show the results of
simulated collisions incorporating the above results to compute the
mean and variance in the event generator of refs. \cite{gp1,gp2}; see
these refs.\ for details. The rapidity densities of baryons,
antibaryons and charged hadrons are taken to be 60, 15 and 300 for
impact parameter $b=0$ and scale with the number of participants for
$b>0$, as appropriate at SPS energy. A value of $f$ is assigned to
each event using the ad hoc distribution, $f(b) =
0.25~[1-(b/b_{0})^{2}]$ for $b < b_{0} = 3$~fm.  The mean baryon
number and its fluctuations are computed at $y\approx \eta=0$ for
$\tau_Q = 5$~fm, $\tau_F = 10$~fm, $D = 6$~fm, and rapidity density
contrasts $\delta N= N_q-N_h = 20$ and 40, corresponding to $\rho_q
-\rho_h\sim 0.10$ and 0.2~fm$^{-3}$ on the scale of normal nuclear
matter density.  The `hadron' curve is computed assuming no
enhancement. The difference between this curve and unity is due to
impact parameter fluctuations (volume and thermal fluctuations
\cite{gp1,gp2} are omitted). The top curve is computed with $\delta N
= 40$ but without diffusion. We see that diffusion is a small effect
for our parameter choices.
\begin{figure} 
\epsfxsize=3.25in
\centerline{\epsffile{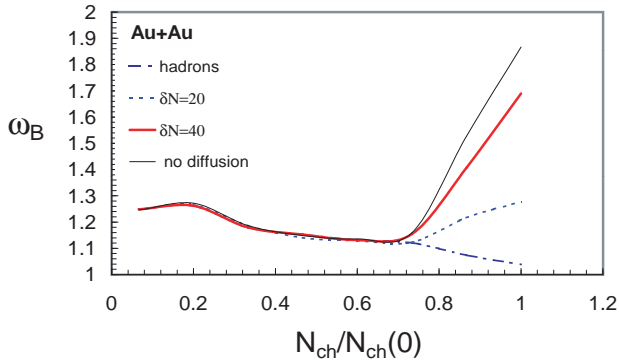}}
\vskip +0.1in
\caption[]{
  Variance (\ref{eq:eq17}) as a function of charged particle multiplicity 
  normalized to the average value at $b=0$.

}\end{figure}

Much more work is needed to obtain crucial information, such as the
beam energies and rapidity ranges needed to probe the high density phase.  
To start, we need a better idea of the phase diagram away from the 
tricritical point.  We must also understand transport processes in the 
unstable and mixed phase regions.  Then, we can combine these ingredients 
to perform dynamical simulations such as those in \cite{spinodal} to assess 
the likelihood of a quench.  We have also overlooked the possibility of 
supeconductivity in the high density phase, a
reasonable assumption near the tricritical point \cite{BergesRajagopal}.  
Superconductivity would modify baryon transport because a) the condensate carries 
baryon current and b) additional transport modes are possible.

In summary, we have explored the possibility that the phase transition
of refs.~\cite{BergesRajagopal,Steph1} can enhance fluctuations of the
net baryon number.  We find that fluctuations can plausibly persist
through freezeout to surmount impact-parameter fluctuations
\cite{BaymHeiselberg}.  A collision that reaches the tricritical point
will have the largest fluctuations, but collisions that pass below can
also be extraordinary, provided that the dynamics can quench the
system.  Calculations for $y\approx 0$ at RHIC energy \cite{gp2}
suggest that proton fluctuations alone can reveal net baryon
fluctuations at the level of fig.~2.  Work is in progress to evaluate
this signal at higher rapidity and lower beam energy where higher
baryon densities are met.

I am grateful to C. Pruneau for many illuminating discussions, and to
J. Berges, K. Rajagopal, E.  Surdutovich and S.  Voloshin for
useful comments. I also thank the Aspen Center for Physics for
hospitality during the completion of this work. This work is supported
in part by the U.S. DOE grant DE-FG02-92ER40713.

\end{narrowtext}

\begin{references}
\bibitem{linear}  Linear contributions to the variance vanish
  during the onset of spinodal decomposition on averaging over
  fluctuations.  At later times, fluctuations are less important,
  so that this term can contribute.
\end{references}
\end{document}